\documentclass{IEEEtran}

\usepackage{graphicx}
\usepackage{amsmath,amssymb,amsthm}
\usepackage{mathtools}
\usepackage{float}
\usepackage{xcolor,colortbl}
\usepackage{color}
\usepackage{fp}
\usepackage{tikz}
\usetikzlibrary{shapes.misc,shadows}
\usepackage{multirow}
\usepackage{xcolor,colortbl}
\usepackage{caption}
\usepackage{graphicx}
\usepackage[linesnumbered,ruled]{algorithm2e}
\usepackage{epstopdf}
\usepackage{pgfplots}
\pgfplotsset{compat=1.7}
\usepackage{caption}
\usepackage{subcaption}

\usepackage{url}
\usepackage[noadjust]{cite}
\usepackage{setspace}
\usepackage{lipsum}
 \usepackage[compact]{titlesec}
    \titlespacing{\section}{0pt}{0.5ex}{0.5ex}
    \titlespacing{\subsection}{0pt}{1ex}{0ex}
    \titlespacing{\subsubsection}{0pt}{0.5ex}{0ex}
\usepackage[abbreviations, automake]{glossaries-extra}
\makeglossaries 
\newabbreviation{OMA}{OMA}{orthogonal multiple access}
\newabbreviation{NOMA}{NOMA}{Non-orthogonal multiple access}
\newabbreviation{SINR}{{SINR}}{signal-to-interference-plus-noise ratio}
\newabbreviation{SIC}{{SIC}}{successive interference cancellation}
\newabbreviation{MSD}{{MSD}}{Minimum SINR  Difference}
\newabbreviation{CDF}{CDF}{cumulative distribution function}
\newabbreviation{BS}{BS}{base station}

\newacronym{N}{\ensuremath{N}}{Number of users connected to the base station}
\newacronym{GammaOMA}{\ensuremath{\gamma_i^\text{\tiny OMA}}}{SINR of the $i^{\text{th}}$ user in OMA system}
\newacronym{GammaNOMA}{\ensuremath{\gamma_i^\text{\tiny NOMA}}}{SINR of the $i^{\text{th}}$ user in NOMA system}
\newacronym{hi}{\ensuremath{h_i}}{Channel gain of the $i^{\text{th}}$ user}
\newacronym{N0}{\ensuremath{\sigma^2}}{Noise variance}
\newacronym{G}{\ensuremath{G}}{Number of users allocated with bandwidth}
\newacronym{I}{\ensuremath{I}}{Inter cell interference}
\newacronym{beta}{\ensuremath{\beta}}{Imperfect SIC parameter}
\newacronym{Pt}{\ensuremath{P_t}}{Available transmit power at the base station}
\newacronym{alphai}{\ensuremath{\alpha_i}}{Fraction of power allocated to the user $i$}
\newacronym{ROMA}{\ensuremath{R_i^\text{\tiny OMA}}}{Normalized downlink rate for a user $i$ in case of OMA system}
\newacronym{RNOMA}{\ensuremath{R_i^\text{\tiny NOMA}}}{Normalized downlink rate for a user $i$ in case of NOMA system}
\newacronym{Ffun}{\ensuremath{\chi_\text{\tiny G}}}{?}
\newacronym{Delta}{\ensuremath{\delta_i}}{?}
\newacronym{DeltaMSD}{\ensuremath{\Delta^{\tiny MSD}_{\tiny i-1, i}}}{?}
\newacronym{MUP}{MUC}{Multi-User Clustering}
\newacronym{AMUP}{AMUC}{Adaptive Multi-User Clustering}
\newacronym{NF}{NF}{Near-Far}

\begin{document}
\bstctlcite{IEEEexample:BSTcontrol}
\title{  {Adaptive Multi-User Clustering and Power Allocation for NOMA Systems with Imperfect SIC} }
\author{\IEEEauthorblockN{Nemalidinne Siva Mouni, \textit{Student Member, IEEE}, Pavan Reddy M., \textit{Student Member, IEEE}, \\Abhinav Kumar, \textit{Senior Member, IEEE}, and Prabhat K. Upadhyay, \textit{Senior Member, IEEE} \vspace{-3em}} 
\thanks{S. M. Nemalidinne, P. R. Manne, and A. Kumar are with the Department of Electrical Engineering, Indian Institute of Technology Hyderabad, Telangana,
India. (e-mail: \{ee19resch11003, ee14resch11005\}@iith.ac.in, abhinavkumar@ee.iith.ac.in).}\thanks{P. K. Upadhyay is with the Department of Electrical Engineering, Indian Institute of Technology Indore, Madhya Pradesh, India. (e-mail: pkupadhyay@iiti.ac.in).
}\thanks{The research of S. M. Nemalidinne is supported in part by the Prime Minister Research Fellowship. The work of A. Kumar is supported in part by TiHAN Faculty Fellowship. The research of P. K. Upadhyay is supported in part by the project under the Visvesvaraya PhD Scheme of Ministry of Electronics \& Information Technology (MeitY), Government of India, being implemented by Digital India Corporation (formerly Media Lab Asia).}
}
\maketitle
\begin{abstract}
Non-orthogonal multiple access (NOMA) is recognized as a promising radio access technique for the next generation wireless systems. We consider a practical downlink NOMA system with imperfect successive interference cancellation and derive bounds on power allocation factors for a NOMA cluster. We propose a minimum signal-to-interference-plus-noise ratio difference criterion between two successive NOMA users in a NOMA cluster to achieve higher rates than an equivalent orthogonal multiple access system.  We then propose adaptive multi-user clustering and power allocation algorithms for downlink NOMA systems. Through extensive simulations, we show that the proposed algorithms achieve higher rates than the state-of-the-art algorithms.
\end{abstract}
\IEEEpeerreviewmaketitle 
\section{Introduction}
\gls{NOMA} is one of the key technologies for fifth-generation (5G) and beyond 5G cellular networks~\cite{SIG1}. In power-domain \gls{NOMA}, the transmitter multiplexes the symbols intended for multiple users with varying power levels and transmits it in the same time-space-frequency resource. The capacity improvement with the multiplexing of users in \gls{NOMA} is significantly dependent on the channel conditions of the users~\cite{Mouni}. Hence, to achieve higher user data rates, the \gls{BS} must optimally choose the number of users to be clustered together for \gls{NOMA} and suitably allocates the available power to all the users in each cluster~\cite{SIG2}. 
 
The practical \gls{NOMA} systems have imperfections in \gls{SIC} which significantly impact the network throughput performance~\cite{jose}. The imperfections in the \gls{SIC} strongly impact the higher channel gain user rates, and hence, impose conditions where \gls{NOMA} pairing for certain users may not be beneficial for the overall system~\cite{Mouni}. However, most of the existing works like~\cite{UP1, UP2} discuss the user clustering algorithms considering two users in each cluster under a perfect \gls{SIC} scenario. A few works in the literature have considered the imperfect \gls{SIC} while proposing 2-user pairing algorithms for \gls{NOMA}~\cite{Mouni,UP3}. Nevertheless, limited analysis has been performed towards user clustering and power allocation for a generalized number of users in each cluster in the presence of imperfections in \gls{SIC}~\cite{UP4, GenPairing}.   

To the best of our knowledge, this is the first work that considers the \gls{MSD} criterion for user clustering and power allocation for the downlink NOMA systems with a generalized number of users in each cluster and with the imperfect \gls{SIC}. Clustering schemes have been proposed in literature under the assumption that \gls{NOMA} consistently outperforms \gls{OMA}. However, it is not always the case in practice.   

Motivated by the aforementioned details, we present the following key contributions in this letter. 
\begin{itemize}
\item We propose bounds on the power allocation for multi-user \gls{NOMA} clusters to achieve higher data rates than an equivalent \gls{OMA} system in presence of imperfect SIC. 
\item We derive bounds on channel coefficients for user clustering in the presence of imperfect SIC. 
\item Using these derived bounds, we present multi-user clustering (\gls{MUP}) and adaptive multi-user clustering (\gls{AMUP}) algorithms for NOMA systems.
\item We then present a novel power allocation procedure for all the users in a cluster. 
\end{itemize}
In this letter, we have formulated an adaptive clustering criterion for multiple users in a \gls{NOMA} cluster by comparing their \gls{NOMA} rates with corresponding \gls{OMA} rates. The performance of the proposed \gls{AMUP} algorithm is always better than \gls{OMA} even in the presence of imperfections in \gls{SIC} and saturates with an increase in the number of users in a cluster but never degrades, unlike the conventional user-pairing algorithms.


\section{System Model}
\begin{figure}
\centering
\includegraphics[scale=0.28]{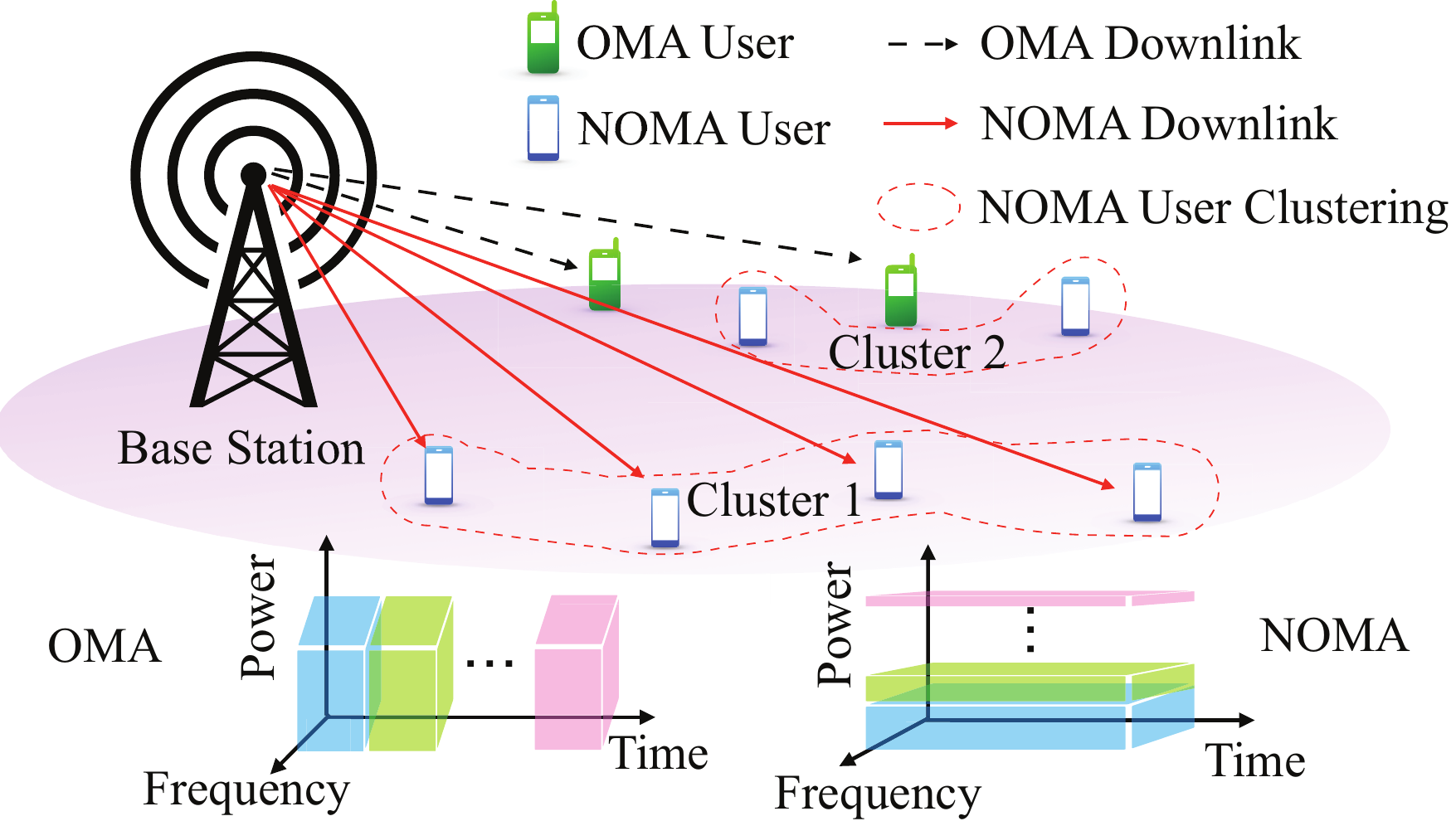}
\caption{System model.}
\label{fig:SystemModel}
\end{figure}
\label{sec:SystemModel}
We consider a set of users $U=\{1, \ldots, \gls{N}\}$, where \gls{N} is the number of users associated to the \gls{BS} under consideration and \gls{G} ($\leq $\gls{N}) users may be clustered as \gls{NOMA} clusters as shown in Fig.~\ref{fig:SystemModel}. The downlink \gls{SINR} for any user $i \in U$ in the case of an \gls{OMA} system is defined as
\begin{align}
\gls{GammaOMA}=\gls{Pt}\dfrac{|\gls{hi}|^2}{\gls{N0}+\gls{I}},
\end{align} 
where \gls{Pt} is the transmit power at the \gls{BS}, \gls{hi} is the channel coefficient between the \gls{BS} and a user $i$, \gls{N0} is the additive white Gaussian noise variance, and \gls{I} is the inter-cell interference. Assuming \gls{G} ($\leq $\gls{N}) users are allocated with the available bandwidth at any time instant, the normalized data rate for a user $i$ in case of an \gls{OMA} system is formulated as 
\begin{align}
\gls{ROMA}=\dfrac{1}{\gls{G}}\log_2(1+\gls{GammaOMA}).\label{eqn:ROMA}
\end{align}

For the \gls{NOMA} formulation, at any time instant, we consider \gls{G} users in a \gls{NOMA} cluster with channel coefficients satisfying $|h_1|^2>|h_2|^2>\ldots>|h_G|^2$. Then, the \gls{SINR} of user $i$ in the \gls{NOMA} system is given by
\begin{align}
\gls{GammaNOMA}&=\dfrac{\gls{alphai}\gls{Pt}|\gls{hi}|^2}{\gls{N0}+\gls{I}+\sum_{j=1, j\neq i}^{i-1}\alpha_j\gls{Pt}|h_i|^2+\gls{beta}\sum_{k=i+1}^{G}\alpha_k\gls{Pt}|h_i|^2}\nonumber\\
&=\dfrac{\gls{alphai}\gls{GammaOMA}}{1+\sum_{j=1, j \neq i}^{i-1}\alpha_j\gls{GammaOMA}+\gls{beta}\sum_{k=i+1}^{G}\alpha_k\gls{GammaOMA}},
\label{eqn:GammaNOMA}\end{align}
where \gls{alphai} is the fraction of total power allocated to the user $i$ and $\gls{beta}\in[0,1]$ is the imperfection in \gls{SIC}~\cite{jose}. Note that $\gls{beta}=0$ indicates a perfect \gls{SIC}. The normalized data rate of the user $i$ in \gls{NOMA} is formulated as 
\begin{align}
\gls{RNOMA}=\log_2(1+\gls{GammaNOMA}).
\label{eqn:RNOMA}
\end{align}
Further, based on \gls{NOMA} principle~\cite{NomaPrinciple}, the following conditions hold.


  \begin{equation}
\alpha_{\text{\scriptsize G}}>\alpha_{\text{\scriptsize G}-1}>... > \alpha_2 > \alpha_1  \label{eqn:IndvAlpha}
  \end{equation}
\begin{equation}
\sum_{i=1}^{G} \alpha_i =1. \hspace{0.4cm} \label{eqn:SumAlpha}
\end{equation}

From \eqref{eqn:IndvAlpha}, we conclude the following result
\begin{align}
\sum_{j=i+1}^{G} \alpha_j&>(G-i) \alpha_i.
\label{eqn:UpperDelta}
\end{align}
Next, we present the bounds on power allocation factors and channel coefficients for users in a \gls{NOMA} cluster in the presence of imperfections in \gls{SIC}.

\vspace{-0.2em}
\section{Computation of Bounds}
\vspace{-0.5em}
\label{sec:Bounds}
In this section, we derive a lower bound on the power allocation factor and an upper bound on the imperfections in \gls{SIC} for a \gls{NOMA} cluster under consideration in terms of channel coefficients. Further, we formulate \gls{MSD} criterion for a \gls{NOMA} multi-user cluster to achieve higher \gls{NOMA} user rates as compared to \gls{OMA}.
\subsection{Lower Bound on \gls{alphai}}
We consider the \gls{NOMA} rate of each individual user should be greater than the \gls{OMA} rate ($\gls{RNOMA}>\gls{ROMA}$). Thus, from \eqref{eqn:ROMA} and \eqref{eqn:RNOMA}, we have
 \begin{align}
\log_2(1+\gls{GammaNOMA})&>\dfrac{1}{\gls{G}}\log_2(1+\gls{GammaOMA}) \nonumber, \textmd{ yielding}\\
\gls{GammaNOMA}&>\left[1+\gls{GammaOMA}\right]^{\frac{1}{\gls{G}}} -1.
\label{eqn:ASR}
\end{align}
\text{Using \gls{GammaNOMA} from \eqref{eqn:GammaNOMA} in \eqref{eqn:ASR}, we have}
\begin{align}
\dfrac{\gls{alphai}\gls{GammaOMA}}{1+\sum_{j=1, j \neq i}^{i-1}\alpha_j\gls{GammaOMA}+\gls{beta}\sum_{j=i+1}^{\gls{G}}\alpha_j\gls{GammaOMA}}&>\left[1+\gls{GammaOMA}\right]^{\frac{1}{\gls{G}}} -1. \nonumber
\end{align}
\text{Substituting} $\sum_{j=1, j \neq i}^{i-1}\alpha_j=1-\alpha_i-\sum_{j=i+1}^{G}\alpha_j $ from \eqref{eqn:SumAlpha}, we get
\begin{align}
\dfrac{\gls{alphai}}{1+\Big[1+(\beta-1)\sum_{j=i+1}^{\gls{G}}\alpha_j-\alpha_i\Big]\gls{GammaOMA}}>
\dfrac{\left[1+\gls{GammaOMA}\right]^{\frac{1}{\gls{G}}} -1}{\gls{GammaOMA}}.
\nonumber
\nonumber
\end{align}
Solving it further, we obtain lower bound on \gls{alphai} as follows:
\begin{align}
\gls{alphai}>\Big({1+\big(1+(\beta-1)\sum_{j=i+1}^{\gls{G}}\alpha_j\big)\gls{GammaOMA}}\Big)\Bigg(\dfrac{(1+\gls{GammaOMA})^\frac{1}{G}-1}{\gls{GammaOMA}(1+\gls{GammaOMA})^\frac{1}{\gls{G}}}\Bigg).
\label{eqn:LowerBound1}
\end{align}
For the ease of understanding, we define
\begin{align}
\gls{Ffun}(\gls{GammaOMA})=&\dfrac{(1+\gls{GammaOMA})^\frac{1}{G}-1}{\gls{GammaOMA}(1+\gls{GammaOMA})^\frac{1}{\gls{G}}}.\label{eqn:Ffun}
\end{align}
Using~\eqref{eqn:Ffun}, we reformulate~\eqref{eqn:LowerBound1} as $\gls{alphai}>\gls{Delta}$, where
\begin{align}
\gls{Delta}=&\Big({1+\big(1+(\beta-1)\sum_{j=i+1}^{\gls{G}}\alpha_j\big)\gls{GammaOMA}}\Big)\gls{Ffun}(\gls{GammaOMA}).
\label{eqn:Delta}
\end{align}
Substituting \eqref{eqn:UpperDelta} in \eqref{eqn:Delta}, we get
\begin{align}
\gls{Delta}<\Big({1+\Big(1+(\beta-1) (G-i)\alpha_i\Big)\gls{GammaOMA}}\Big)\gls{Ffun}(\gls{GammaOMA}).
\end{align}
Thus, if $\gls{alphai}>\Big({1+\Big(1+(\beta-1) (G-i)\alpha_i\Big)\gls{GammaOMA}}\Big)\gls{Ffun}(\gls{GammaOMA})$, then, $\gls{alphai}>\gls{Delta}$. Solving $\gls{alphai}$ further, we obtain
\begin{align}
\gls{alphai}&>\dfrac{(1+\gls{GammaOMA})\gls{Ffun}(\gls{GammaOMA})}{1-(\gls{beta}-1)(G-i)\gls{GammaOMA}\gls{Ffun}(\gls{GammaOMA})}.\label{eqn:LowerBound2}
\end{align}
Note that if \eqref{eqn:LowerBound2} is satisfied, then \gls{RNOMA} is always greater than \gls{ROMA}. Further, \eqref{eqn:LowerBound1} is a sufficient condition to achieve higher \gls{NOMA} rates,  whereas \eqref{eqn:LowerBound2} is a much stricter bound as compared to \eqref{eqn:LowerBound1}. 
Since \eqref{eqn:LowerBound2} is dependent only on \gls{GammaOMA} and \gls{G}, we use \eqref{eqn:LowerBound2} to define  constraints for multi-user clustering. 
\vspace{-0.5em}
\subsection{Upper Bound on the Imperfect SIC Parameter ($\gls{beta}$)}
\vspace{-0.3em}
From \eqref{eqn:IndvAlpha}, $\alpha_i>\alpha_{i-1}$. Using the bounds in \eqref{eqn:LowerBound2}, we get

{
\begin{align}
\dfrac{(1+\gamma_{i}^{\text{\tiny OMA}})\gls{Ffun}(\gamma_{i}^{\text{\tiny OMA}})}{1-(\gls{beta}-1)(G-i)\gamma_{i}^{\text{\tiny OMA}}\gls{Ffun}(\gamma_{i}^{\text{\tiny OMA}})}\hspace{-3cm}&\nonumber \\
&>
\dfrac{(1+\gamma_{i-1}^{\text{\tiny OMA}})\gls{Ffun}(\gamma_{i-1}^{\text{\tiny OMA}})}{1-(\gls{beta}-1)(G-i+1)\gamma_{i-1}^{\text{\tiny OMA}}\gls{Ffun}(\gamma_{i-1}^{\text{\tiny OMA}})}.
\label{eqn:Alpha}
\end{align}}
For ease of exposition, we define the following terms
\begin{align}
D_i&=\dfrac{(1+\gamma_{i-1}^{\text{\tiny OMA}})\gls{Ffun}(\gamma_{i-1}^{\text{\tiny OMA}})}{(1+\gamma_{i}^{\text{\tiny OMA}})\gls{Ffun}(\gamma_{i}^{\text{\tiny OMA}})},\nonumber
E_{i}=(G-i)\gamma_i^{\text{\tiny OMA}}\gls{Ffun}(\gamma_i^{\text{\tiny OMA}}),\\
E_{i-1}&=(G-i+1)\gamma_{i-1}^{\text{\tiny OMA}}\gls{Ffun}(\gamma_{i-1}^{\text{\tiny OMA}}).\label{eqn:Def3} \nonumber
\end{align}
Substituting $D_i$, $E_{i}$ and $E_{i-1}$ in \eqref{eqn:Alpha}, we get 
\begin{align}
1-(\gls{beta}-1)E_{i-1}&>D_i\big[1-(\gls{beta}-1)E_i\big], \\
\gls{beta}&<\dfrac{1-D_i}{E_{i-1}-D_i E_{i}}+1\triangleq \zeta_{i-1,i}.\label{eqn:beta}
\end{align}
Thus, for the \gls{NOMA} rates to be higher than the \gls{OMA} rates, the typical value of imperfections in \gls{SIC} must satisfy the constraint in \eqref{eqn:beta} for the multi-user clustering in \gls{NOMA} under consideration.
\subsection{\gls{MSD} between Successive Users}
In case of $\gls{G}$ users clustered in \gls{NOMA} system, we define the \gls{MSD} between two users for achieving higher \gls{NOMA} rates as follows. We apply positivity constraint $\gls{beta}>0$ in \eqref{eqn:beta}, thus,
\begin{align}
1-D_i >D_i E_i-E_{i-1} \rightarrow E_{i-1} >D_i E_{i}+D_i -1 .\label{eqn:Def4}
\end{align}
Substituting $E_{i-1}$ from \eqref{eqn:Def3} in \eqref{eqn:Def4}, we get
\begin{align}
(G-i+1)\gamma_{i-1}^{\text{\tiny OMA}}\gls{Ffun}(\gamma_{i-1}^{\text{\tiny OMA}})&>D_i E_i+D_i -1,\nonumber\\
\gamma_{i-1}^{\text{\tiny OMA}}&>\dfrac{D_i E_i+D_i-1}{(G-i+1)\gls{Ffun}(\gamma_{i-1}^{\text{\tiny OMA}})},\nonumber\\
\gamma_{i-1}^{\text{\tiny OMA}}-\gamma_{i}^{\text{\tiny OMA}}&>\dfrac{D_i E_i+D_i -1}{(G-i+1)\gls{Ffun}(\gamma_{i-1}^{\text{\tiny OMA}})}-\gamma_{i}^{\text{\tiny OMA}}.\nonumber
\end{align}
Thus, we define the \gls{MSD} between users $i-1$ and $i$ as follows
\begin{align}
\gls{DeltaMSD}=\dfrac{D_i E_i+D_i -1}{(G-i+1)\gls{Ffun}(\gamma_{i-1}^{\text{\tiny OMA}})}-\gamma_{i}^{\text{\tiny OMA}}.	
\label{eqn:MSD}
\end{align}
Note that for a cluster of $G$ users, we have $G-1$ combinations of \gls{DeltaMSD} and $\zeta_{i-1,i}$ values. 
In case \eqref{eqn:beta} and \eqref{eqn:MSD} are satisfied for all these combinations, then all the $G$ users can be clustered together to achieve \gls{NOMA} rates higher than their \gls{OMA} counterparts.
Further, we next use the \gls{MSD} formulated in \eqref{eqn:MSD} and the lower bound on power allocation formulated in \eqref{eqn:LowerBound1} to propose \gls{MUP}, \gls{AMUP} and power allocation algorithms for \gls{NOMA} systems, respectively.
\begin{figure}[t]
\centering
\includegraphics[scale=0.1]{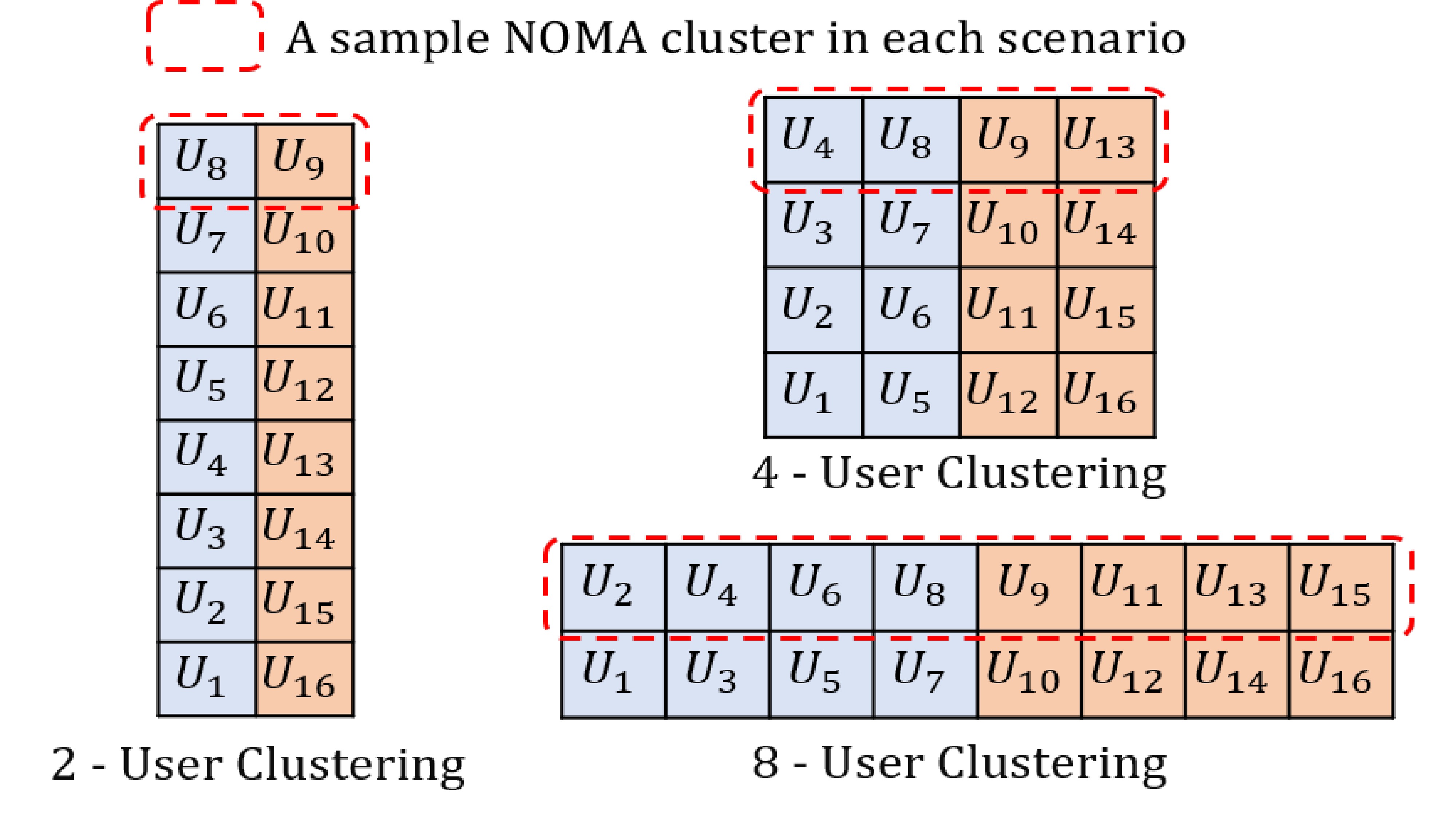}
\caption{Various user clustering scenarios for multi-user \gls{NOMA}.}
\label{fig:Grouping}
\end{figure}
\section{Proposed Algorithms}
\label{sec:Proposed}
In this section, we explain the procedure for the clustering of users. We propose \gls{MUP} and \gls{AMUP} algorithms to achieve higher \gls{NOMA} user rates and compare their performance with a conventional near-far (NF)~\cite{NearFar2} user pairing algorithm. Given a set of users in a cluster, we then present power allocation for each user.

\begin{figure}[t]
\centering
\includegraphics[scale=0.52]{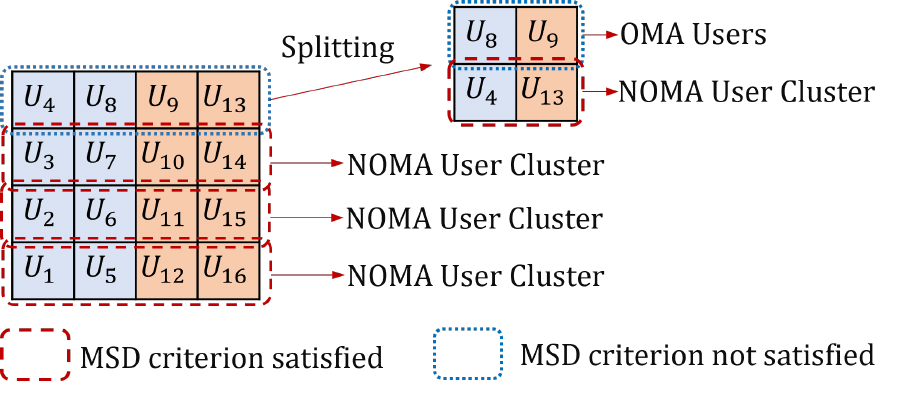}
\caption{Splitting of a 4 user cluster into two 2 user clusters.}
\label{fig:Splitting}
\end{figure}

\SetAlCapNameFnt{\small}
\SetAlCapFnt{\small}
\begin{algorithm}[t]  
\small
\setstretch{0.9}
\SetKwInOut{Input}{Input}
\SetKwInOut{Var}{Variables}
\Input{$U$, $\gamma_\text{i}^\text{\tiny OMA}$, $N$, $G$, $\beta$.}
\Var{$r$ and $i$ are variables representing cluster index and user index, respectively. \textit{Flag} is a variable used to keep track of \gls{MSD} criterion.}
\BlankLine
\caption{Proposed Algorithms}
\label{algo:proposed} 
 \SetKwFunction{FMain}{Clustering}
  \SetKwProg{Fn}{Function}{:}{}
  \Fn{\FMain{$U$, $G$, $N$}}{ $U'=\texttt{sort}(U,\texttt{`descend'})$\;
$U''\leftarrow$ Reshape $U'$ into $G$ columns\; 
$U_\text{\tiny \it Clustered}\leftarrow$ Flip users of $U''$ in columns $\frac{G}{2}$ to $G$\;
\For{ $r =1 \rightarrow \frac{N}{G}$}{
$U_r\leftarrow$ Pick $G$ users from $r^{th}$ row of $U_\text{\tiny \it Clustered}$\;
\For{$i=2 \rightarrow G$}
{\eIf{  $\gls{beta} < \zeta_{i-1,i}$  \textmd{ \textbf{and} }  $\gamma_{i-1}^{\text{\tiny OMA}}-\gamma_{i}^{\text{\tiny OMA}} > \gls{DeltaMSD}$  }
{\textit{Flag}=0; \\
\For{$i=G\rightarrow 1$}
{$\gls{alphai}=\Big({1+\big(1-\sum_{j=i+1}^{\gls{G}}\alpha_j\big)\gls{GammaOMA}}\Big)\gls{Ffun}(\gls{GammaOMA}$)
}

\eIf {$\sum_{k=1}^{\gls{G}} \alpha_k \leq 1$}
{
\textit{Flag = 0}\;
}
{\textit{Flag}=1\; \textbf{break};}

}{\textit{Flag}=1\; \textbf{break};}}
\eIf{\textit{Flag}=0}{Consider all the users for \gls{NOMA};}
{
\eIf {\gls{AMUP}}
{
\texttt{Clustering}($U_{r}$, $\frac{G}{2}$, $G$)\;
}
{\textit{\gls{MUP}}\; Consider all the users for \gls{OMA};}}
}}
\end{algorithm}
\vspace{-0.05in}
\subsection{Multi-User Clustering (\gls{MUP}) Algorithm}
We use the upper bound on \gls{beta} \eqref{eqn:beta} and the \gls{MSD} criterion \eqref{eqn:MSD} to propose an \gls{MUP} algorithm for a generalized number of users as follows. With \gls{G} users in a cluster, we first evaluate $\zeta_{i-1,i}$ as in \eqref{eqn:beta} and \gls{DeltaMSD} as in \eqref{eqn:MSD} for all the $\gls{G}-1$ combinations. We consider clustering these \gls{G} users in \gls{NOMA} only when \eqref{eqn:beta} and \eqref{eqn:MSD} are satisfied for all the $\gls{G}-1$ combinations. Else, all the \gls{G} users are designated as \gls{OMA} users. This way, the individual rates of each user will never be less than that of the corresponding \gls{OMA} rates. Since, we check the criteria in \eqref{eqn:beta} and \eqref{eqn:MSD} for $G-1$ times in each of $N/G$ clusters, the complexity of the proposed \gls{MUP} is $\mathcal{O}\big(2N /G \times(G-1)\big)$. Further, the probability of users not meeting \gls{MSD} criterion increases with an increase in the number of users in a cluster and imperfections in \gls{SIC}. Hence, with large sized user clustering and higher values of $\beta$, the \gls{MUP} algorithm will designate most of the users as \gls{OMA}. To address this issue, next, we propose an \gls{AMUP} algorithm.
\vspace{-0.05 cm}
\subsection{Adaptive Multi-User Clustering (\gls{AMUP}) Algorithm}
In the \gls{AMUP} algorithm, whenever the \gls{G} users in a cluster fail to meet $\zeta_{i-1,i}$ and \gls{DeltaMSD}, we split the cluster into two halves. For this new cluster of users, we again perform the criterion check as formulated in \eqref{eqn:beta} and \eqref{eqn:MSD}. If the criterion is met for all the combinations, we continue clustering those $\gls{G}/2$ users in \gls{NOMA}. Otherwise, we continue splitting this cluster again into two new halves. We continue this procedure until we end up with a single user. If the \gls{MSD} criterion is not met for any user clustering, then that single user will be designated as \gls{OMA} user. Further, while splitting a cluster into two halves, we follow the \gls{NF}~\cite{NearFar2} user pairing procedure. Otherwise, the throughput gains will not be achieved. In Fig.~\ref{fig:Splitting}, we have presented an example of splitting a 4-user cluster into two 2-user clusters. Note that the 2 users clustered together after the splitting process in Fig.~\ref{fig:Splitting} are exactly the same as they would have been in the case of 2-user clustering in Fig.~\ref{fig:Grouping}. 

The complexity of the proposed algorithm is calculated as follows. For each of $N/G$ clusters, the proposed algorithm initially checks \eqref{eqn:beta} and \eqref{eqn:MSD} criterion for $(G-1)$ combinations. If the conditions are not satisfied, then it splits the cluster into two halves and checks the criterions again. This results into additionally $4\times((G/2) -1)$ computations. This procedure is continued till only one user is left in a cluster. Thus, in a worst case scenario, the complexity of the proposed \gls{AMUP} algorithm is $\mathcal{O}\Big((2N/G)  \big(G\log_2G-((1-G)/(1-\log_2G))\big)\Big)$. Next, we propose the power allocation algorithm for a given set of clustered users.

\begin{figure}[t!]
\centering
\includegraphics[scale=0.38]{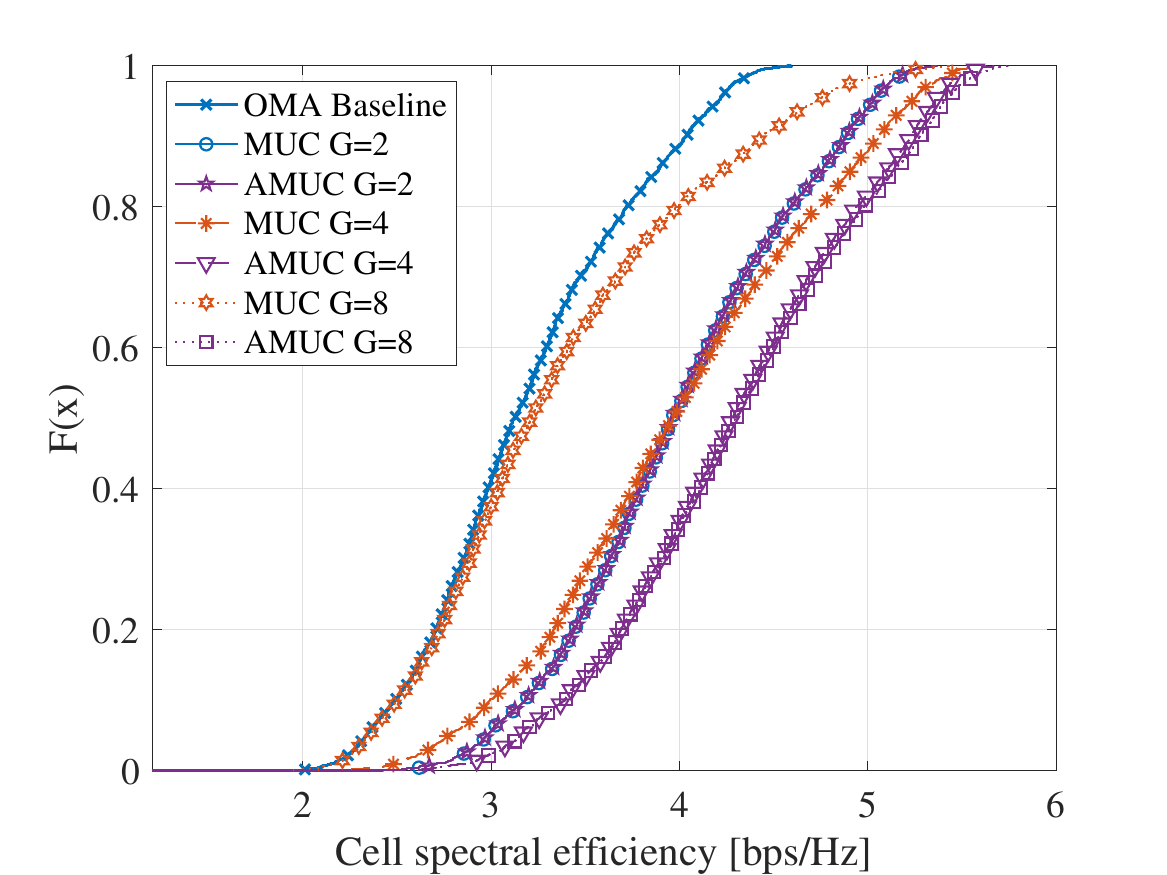}
\caption{CDF of cell spectral efficiency with varying number of users in each cluster and perfect \gls{SIC} $(\gls{beta}=0)$.}
\label{fig:Beta0C}
\end{figure}
\vspace{-0.05cm}
\subsection{Power Allocation}
Based on the lower bound formulated in \eqref{eqn:LowerBound1}, we allocate the minimum power required for each user as follows
\begin{align}
\gls{alphai}=&\Big({1+\big(1-\sum_{j=i+1}^{\gls{G}}\alpha_j\big)\gls{GammaOMA}}\Big)\gls{Ffun}(\gls{GammaOMA}).
\nonumber
\end{align}
We begin allocation with user $G$, as $\alpha_\text{\tiny G}$ is dependent only on $\gamma^\text{\tiny OMA}_\text{\tiny G}$. We then use this
allocated power $\alpha_\text{\tiny G}$ and  $\gamma^\text{\tiny OMA}_{\text{\tiny G}-1}$ to recursively compute the power allocation factor $\alpha_{\text{\tiny G}-1}$. Likewise, we continue allocation till $\alpha_\text{\tiny 1}$. Further, when we allocate the minimum power required for each user based on \eqref{eqn:LowerBound1}, $\sum_{i=1}^{G}\alpha_i$ is less than 1. Hence, to maximize the achievable sum rates, we allocate the remaining power $\big(1-\sum_{i=1}^G\big)\gls{Pt}$ to the strong user. 

We have presented a pseudo-code to implement the proposed \gls{MUP}, \gls{AMUP}, and power allocation in Algorithm~\ref{algo:proposed}. Next, we present the simulation results.

\section{Results and Discussion}
\label{sec:Results}

\begin{figure}[t]
\centering
\includegraphics[scale=0.38]{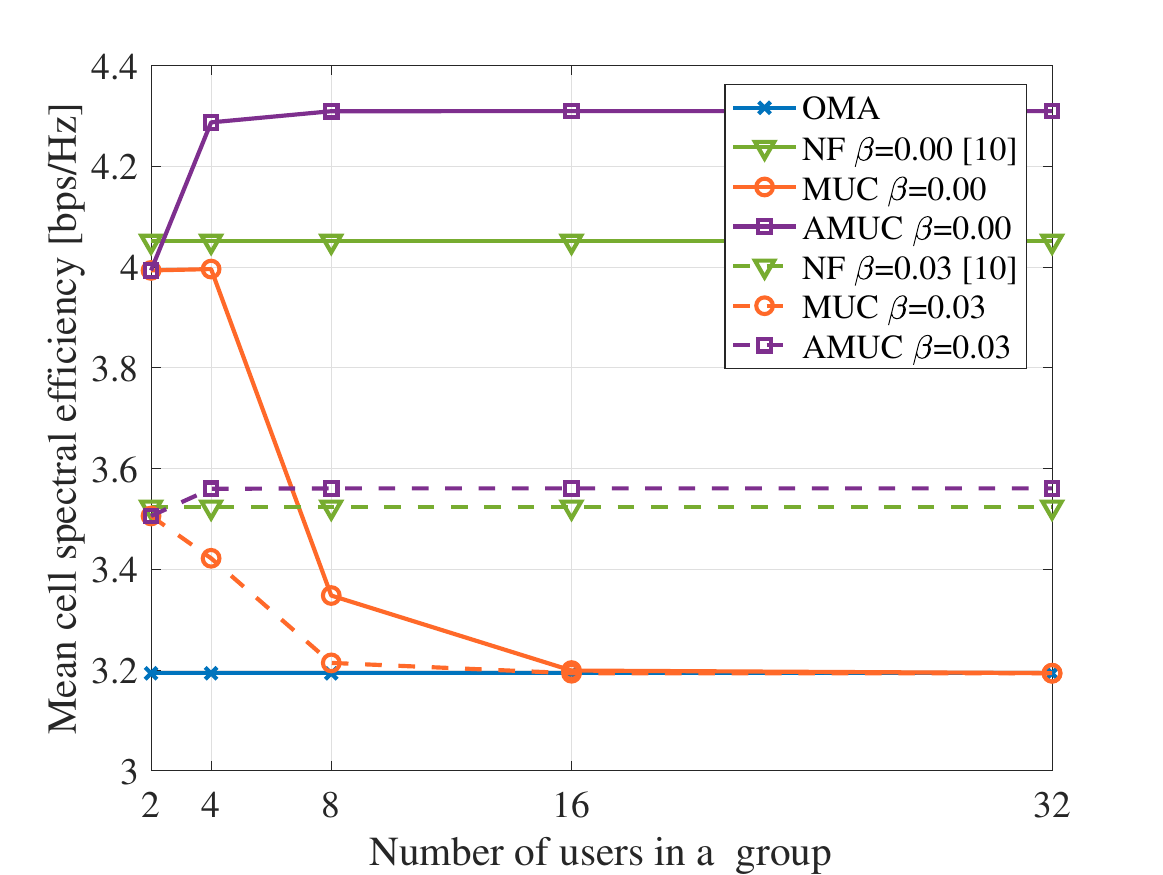}
\caption{Mean cell spectral efficiency comparison with varying number of users in each cluster.}
\label{fig:Beta0M}
\end{figure}

For the evaluation of the proposed algorithms, we have considered Poisson point distributed \gls{BS}s and users with densities 25 BS/km$^2$ and 2000 users/km$^2$, respectively.  The simulation parameters and the path loss model considered for the evaluation are as per the urban cellular scenario presented in \cite{38901}. For each user, we calculate the received \gls{SINR} from each \gls{BS} and then associate the user to the \gls{BS} from which it receives maximum \gls{SINR}. We then randomly pick $N=64$ users associated with each \gls{BS} and perform evaluation for user clusterings with $G\in\{2,\ldots,32\}$. 
For each $G$ value, we perform the proposed user clustering, \gls{MUP}, \gls{AMUP}, and power allocation as mentioned in Algorithm~\ref{algo:proposed}. We then perform Monte-Carlo simulations to calculate the cell spectral efficiency for each algorithm.

In Fig.~\ref{fig:Beta0C}, we plot \gls{CDF} of cell spectral efficiency for varying number of users in each cluster and for perfect \gls{SIC} $(\gls{beta}=0)$. The performance of the \gls{OMA} system does not vary with number of users in each cluster.  With \gls{MUP}, \gls{NOMA} achieves higher performance than \gls{OMA}, but the performance degrades as we further increase the number of users in each cluster. This is because, with a larger number of users in a cluster, the constraint \eqref{eqn:MSD} fails more often, and all the users in the cluster will be designated as \gls{OMA}. Thus, the performance degrades with an increase in user clustering, but it still outperforms the \gls{OMA}. With \gls{AMUP}, whenever the constraint \eqref{eqn:MSD} fails, the algorithm tries to cluster the users in a smaller size. Hence, the performance saturates but never degrades with an increase in the number of users in a cluster. A similar trend is observed in Fig.~\ref{fig:Beta0M}, where we have presented the variation of mean cell-spectral efficiency with various clustering algorithms. Note that we have evaluated the conventional \gls{NF} user pairing~\cite{NearFar2}. This algorithm always considers only two users in a cluster, and hence, the mean cell spectral efficiency is constant for any number of users in a cluster. As shown in Fig.~\ref{fig:Beta0M}, with large user clustering ($G>8$), the proposed \gls{AMUP} algorithm performs 37.2\%, and 8.3\% better than the \gls{OMA} baseline, and the conventional \gls{NF} algorithm~\cite{NearFar2} in terms of mean cell spectral efficiency for $\beta = 0$, respectively. 
\begin{figure}[t]
\centering
\includegraphics[scale=0.38]{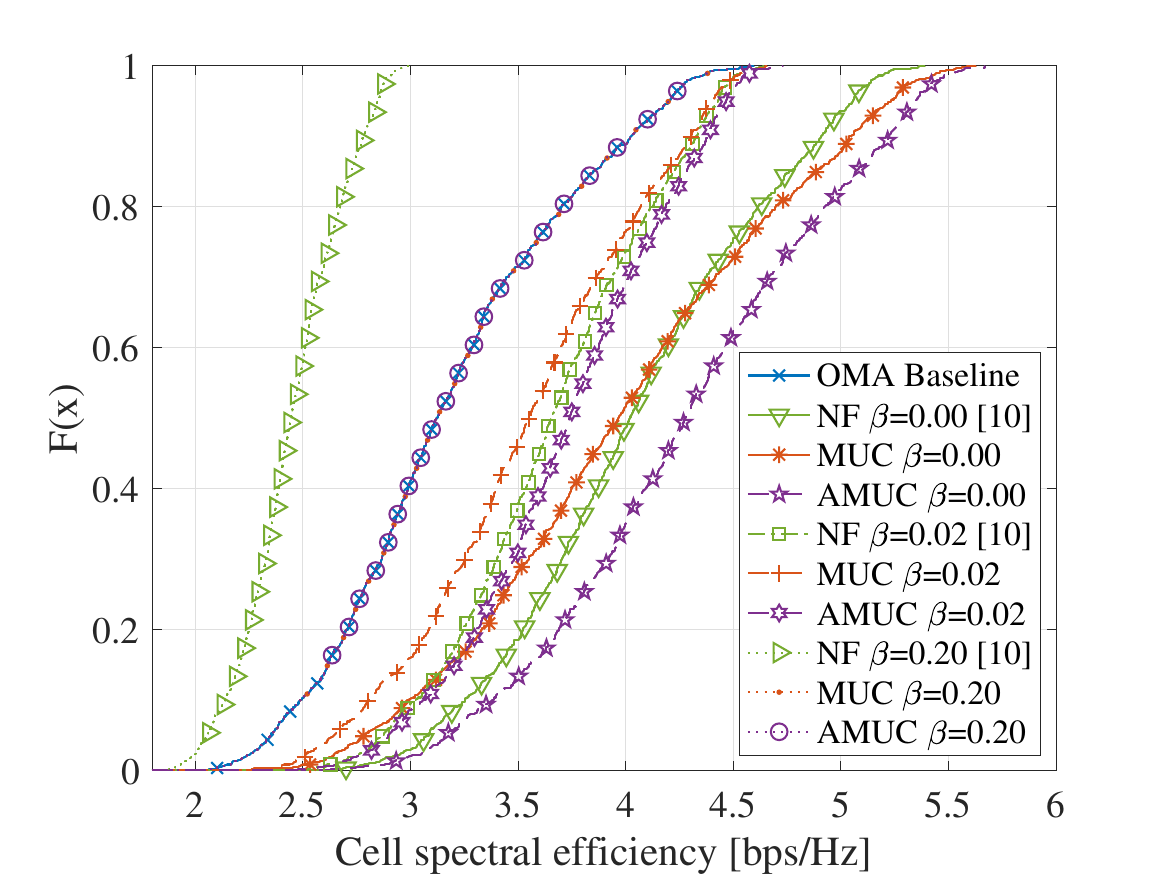}
\caption{CDF of cell spectral efficiency with 4 users in each cluster and varying imperfection in  \gls{SIC} $(\gls{beta})$.}
\label{fig:BetaC}
\end{figure}

In Fig.~\ref{fig:BetaC}, we plot \gls{CDF} of cell spectral efficiency with 4 users in each cluster and varying imperfection in \gls{SIC}. The performance of \gls{NOMA} systems decreases with an increase in \gls{beta}. Whenever the user channel coefficients do not satisfy the constraint in~\eqref{eqn:beta} and~\eqref{eqn:MSD}, the performance with \gls{NOMA} systems is equivalent to \gls{OMA} systems (here for 4-user clustering, bound on $\gls{beta}$ is observed to be $0.0561$). However, the \gls{CDF} of cell spectral efficiency of \gls{NF} algorithm decreases beyond \gls{OMA} for higher values of imperfections in \gls{SIC}.

\begin{figure}[t]
\centering
\includegraphics[scale=0.38]{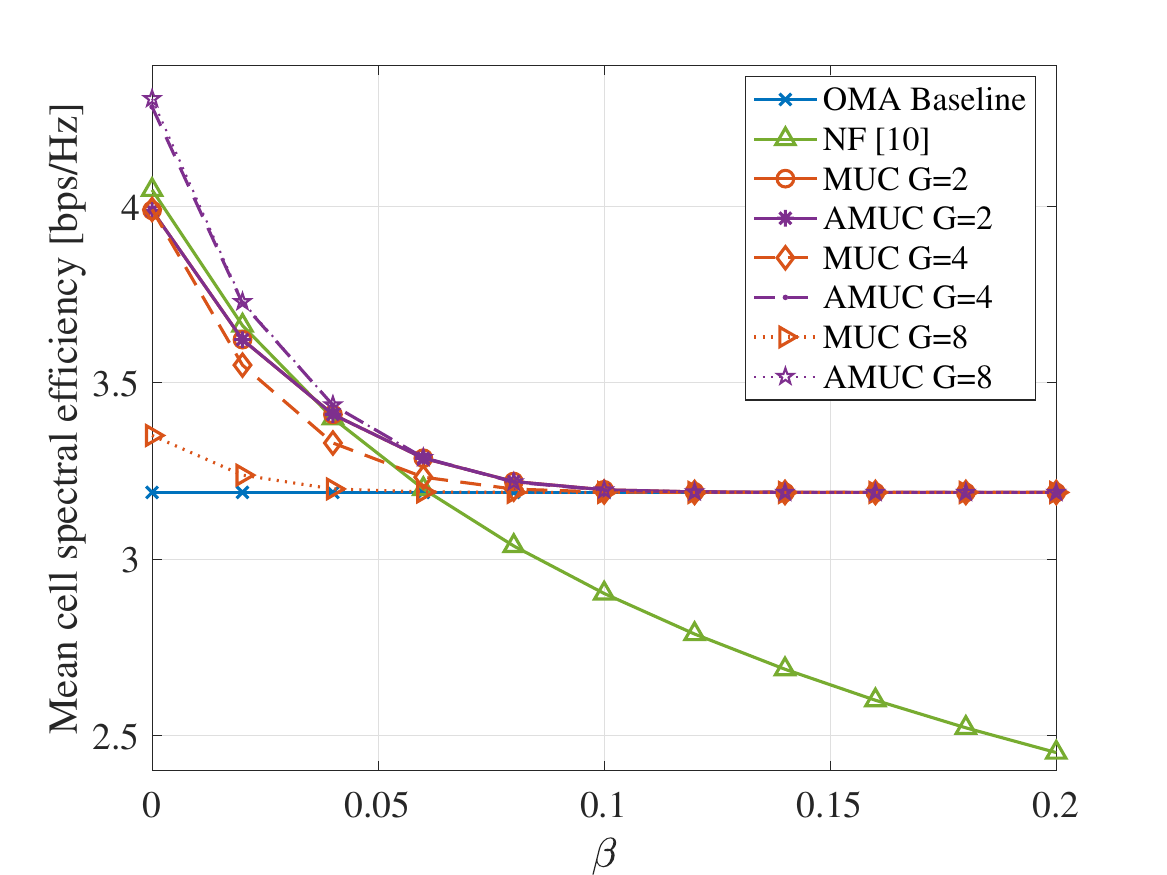}
\caption{Mean cell spectral efficiency comparison with varying imperfection in  \gls{SIC} $(\gls{beta})$.}
\label{fig:BetaM}
\end{figure}

 In Fig.~\ref{fig:BetaM}, we present the mean cell spectral efficiency with varying imperfection in \gls{SIC}. The performance of \gls{OMA} does not change with number of users in each cluster and varying \gls{beta}. The mean cell spectral efficiency of \gls{NF}-based algorithm decreases beyond \gls{OMA} with increasing values of $\beta$. In case of \gls{MUP}, for increasing $\gls{beta}$ values, the mean cell spectral efficiency decreases gradually and converges with \gls{OMA}. Further, for smaller value of $\beta$, the mean cell spectral efficiency decreases with increasing number of users per cluster. With larger user clustering, the performance of \gls{MUP} will be close to \gls{OMA} systems as the condition fails most of the time. However, for smaller $\gls{beta}$, the performance of the proposed \gls{AMUP} saturates with an increase in user clustering, but it does not degrade.  For larger $\gls{beta}$, when the channel coefficients do not satisfy the upper bound on \gls{beta} presented in \eqref{eqn:beta}, the performance converges with \gls{OMA}. In such scenarios, \gls{AMUP} tries to avoid clustering, and hence, convergence with \gls{OMA}. Thus, in the presence of imperfect \gls{SIC}, \gls{NOMA} has superior performance only when the user's channel coefficients satisfy~\eqref{eqn:beta} and~\eqref{eqn:MSD}. 
\section{Conclusion}
\label{sec:Conclusion}
We have derived bounds on power allocation, imperfect \gls{SIC}, and channel coefficients for multi-user clustering in a downlink \gls{NOMA} system. We have proposed adaptive multi-user clustering and power allocation algorithms for a generalized number of users in a cluster. We have derived conditions and shown that it is beneficial to not utilize \gls{NOMA} for certain users to achieve higher user data rates. We have also shown that an increase in the number of users in a cluster does not always achieve higher \gls{NOMA} rates.   In future, we plan to implement and evaluate the proposed algorithms on hardware testbeds.
\bibliographystyle{ieeetran}
\bibliography{Bibfile.bib}

\begin{thebibliography}{10}
\providecommand{\url}[1]{#1}
\csname url@samestyle\endcsname
\providecommand{\newblock}{\relax}
\providecommand{\bibinfo}[2]{#2}
\providecommand{\BIBentrySTDinterwordspacing}{\spaceskip=0pt\relax}
\providecommand{\BIBentryALTinterwordstretchfactor}{4}
\providecommand{\BIBentryALTinterwordspacing}{\spaceskip=\fontdimen2\font plus
\BIBentryALTinterwordstretchfactor\fontdimen3\font minus
  \fontdimen4\font\relax}
\providecommand{\BIBforeignlanguage}[2]{{%
\expandafter\ifx\csname l@#1\endcsname\relax
\typeout{** WARNING: IEEEtran.bst: No hyphenation pattern has been}%
\typeout{** loaded for the language `#1'. Using the pattern for}%
\typeout{** the default language instead.}%
\else
\language=\csname l@#1\endcsname
\fi
#2}}
\providecommand{\BIBdecl}{\relax}
\BIBdecl

\bibitem{SIG1}
Z.~Ding \emph{et~al.}, ``On the performance of non-orthogonal multiple access
  in {5G} systems with randomly deployed users,'' \emph{IEEE Signal Process.
  Lett.}, vol.~21, no.~12, pp. 1501--1505, 2014.

\bibitem{Mouni}
N.~S. Mouni \emph{et~al.}, ``Adaptive user pairing for {NOMA} systems with
  imperfect {SIC},'' \emph{IEEE Wireless Commun. Lett.}, vol.~10, no.~7, pp.
  1547--1551, 2021.

\bibitem{SIG2}
J.~Cui \emph{et~al.}, ``A novel power allocation scheme under outage
  constraints in {NOMA} systems,'' \emph{IEEE Signal Process. Lett.}, vol.~23,
  no.~9, pp. 1226--1230, 2016.

\bibitem{jose}
J.~Jose \emph{et~al.}, ``{VFD-NOMA} under imperfect {SIC} and residual
  inter-relay interference over generalized nakagami-m fading channels,''
  \emph{IEEE Commun. Lett.}, vol.~25, no.~2, pp. 646--650, 2021.

\bibitem{UP1}
K.-H. Nguyen \emph{et~al.}, ``On the energy efficiency maximization of
  {NOMA}-aided downlink networks with dynamic user pairing,'' \emph{IEEE
  Access}, vol.~10, pp. 35\,131--35\,145, 2022.

\bibitem{UP2}
S.~B. Melhem \emph{et~al.}, ``User pairing and outage analysis in multi-carrier
  {NOMA-THz} networks,'' \emph{IEEE Trans. Veh. Technol.}, vol.~71, no.~5, pp.
  5546--5551, 2022.

\bibitem{UP3}
P.~K. Hota \emph{et~al.}, ``Ergodic performance of downlink untrusted {NOMA}
  system with imperfect {SIC},'' \emph{IEEE Commun. Lett.}, vol.~26, no.~1, pp.
  23--26, 2022.

\bibitem{UP4}
X.~Qi \emph{et~al.}, ``Energy-efficient power allocation in multi-user
  mmwave-{NOMA} systems with finite resolution analog precoding,'' \emph{IEEE
  Trans. Veh. Technol.}, vol.~71, no.~4, pp. 3750--3759, 2022.

\bibitem{GenPairing}
M.~{Zeng} \emph{et~al.}, ``Capacity comparison between {MIMO-NOMA} and
  {MIMO-OMA} with multiple users in a cluster,'' \emph{IEEE J. Sel. Areas
  Commun.}, vol.~35, no.~10, pp. 2413--2424, 2017.

\bibitem{NomaPrinciple}
Y.~{Liu} \emph{et~al.}, ``On the capacity comparison between {MIMO-NOMA} and
  {MIMO-OMA},'' \emph{IEEE Access}, vol.~4, pp. 2123--2129, 2016.

\bibitem{NearFar2}
M.~B. {Shahab} \emph{et~al.}, ``User pairing schemes for capacity maximization
  in non‐orthogonal multiple access systems.'' \emph{Wireless Commun. Mobile
  Comput.}, vol.~16, no.~17, pp. 2884--2894, Dec. 2016.

\bibitem{38901}
{3GPP}, ``{Study on channel model for frequencies from 0.5 to 100 GHz},'' {3rd
  Generation Partnership Project (3GPP)}, Technical Report, 38.901, v~16.1.0,
  Jan.~2020.

\end{thebibliography}

\end{document}